\documentclass{eas}
\usepackage{graphicx}
\def\dg{^{\circ}}
\def\mum{\mu {\rm m}}
\def\simlt{\mathrel{\spose{\lower 3pt\hbox{$\mathchar"218$}}\raise 2.0pt\hbox{$\mathchar"13C$}}}
\def\simgt{\mathrel{\spose{\lower 3pt\hbox{$\mathchar"218$}}\raise 2.0pt\hbox{$\mathchar"13E$}}}
\begin{document}

\title{PILOT: design and capabilities}
\runningtitle{PILOT: design and capabilities}

\author{W. Saunders}\address{Anglo-Australian Observatory, Epping, NSW 1710, Australia; \email{will@aao.gov.au}}
\author{P.R. Gillingham }\sameaddress{1}
\author{A.J. McGrath }\sameaddress{1}
\author{ J.W.V. Storey}\address{School of Physics, University of New South Wales, Sydney, NSW 2052}
\author{J.S. Lawrence}\sameaddress{2}
\begin{abstract}
The proposed design for PILOT is a general-purpose, wide-field ($1\dg$) 2.4m, f/10 Ritchey-Chr\'etien telescope, with fast tip-tilt guiding, for use $0.5-25\mum$. The design allows both wide-field and diffraction-limited use at these wavelengths. The expected overall image quality, including median seeing, is $0.28-0.3''$ FWHM from $0.8-2.4\mum$. Point source sensitivities are estimated.

\end{abstract}
\maketitle

\section{Introduction}

The free atmospheric conditions at Dome C are known to be exceptional. The seeing (above $\sim$ 30m), coherence time, isoplanatic angle, infrared sky emission, water vapour absorption and telescope thermal emission are all better than at the best mid-latitude sites (Lawrence {\em et al.\/} \cite{Lawrence},
Agabi {\em et al.\/} \cite{Agabi}, Walden {\em et al.\/} \cite{Walden}).

PILOT (the Pathfinder for an International Large Optical Telescope) is intended to show that we can fully utilise these conditions for optical/infra-red astronomy. It is also intended to demonstrate that large optical telescopes can be built and operated in Antarctica; to fully characterise the possibilities for adaptive optics; and to perform cutting-edge science in its own right.
A design study by the AAO and UNSW is currently underway, with completion mid-2008. As part of that study, we are actively seeking partners in the PILOT project, and input into the design and the scientific requirements

\begin{figure}
\begin{center}
\includegraphics[width=9.5cm,angle=-90]{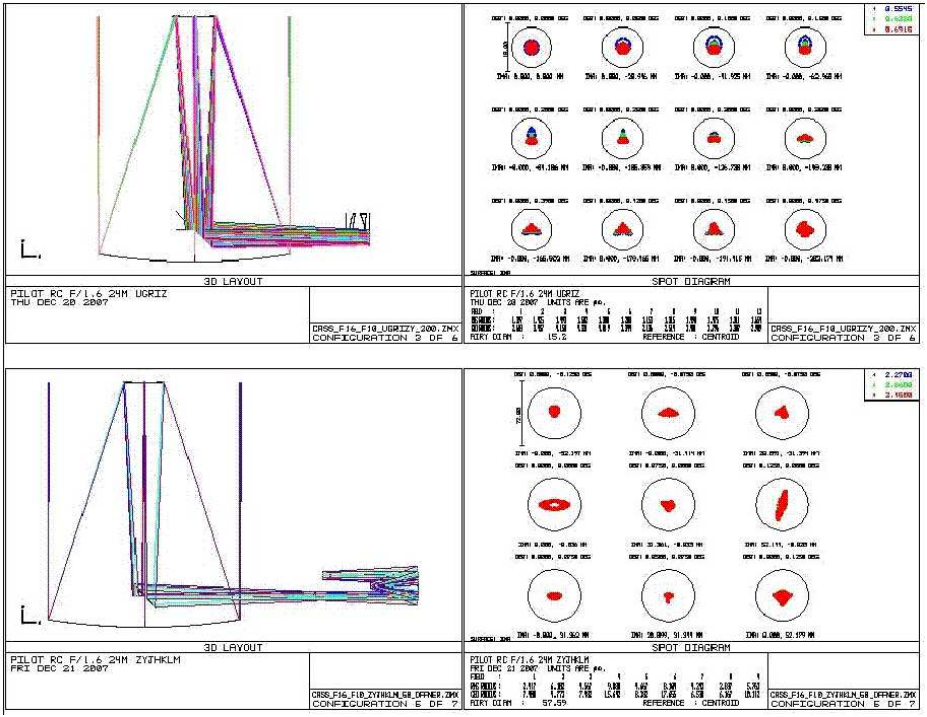}
\end{center}
\caption{Optical layout for PILOT, for wide-field optical ($r$-band) use with $1\dg$ field (above) and wide-field infra-red ($K_{dark}$-band) use with $15'$ field and cold stop (below). The spot diagrams (on right) include the Airy disc and a scale bar of 18$\mum$/$0.15''$ (above) and 72$\mum$/$0.62''$ (below). The optical design is diffraction limited above 500nm.}
\end{figure}

\section{Design overview}
The design has undergone significant evolution during the design study. The current design (December 2007), is for a wide-field (1$\dg$) Ritchey-Chr\'etien system, with 2.4m primary, tip-tilt secondary, corrector lenses for wide-field use,  2 Nasmyth foci, and f/10 overall speed. It will be on a $\sim 30$m tower, and enclosed in a thermally and humidity-controlled dome. All routine operations will be remotely controlled.

A Gregorian design offers a potential advantage for tip-tilt correction, because of the conjugacy of the secondary to residual boundary layer turbulence. However, we calculate that this gain is very small in practice, and far outweighed by the drawbacks of a faster primary, a larger secondary and smaller field of view.

Originally, it was envisaged that Narcissus mirrors would be used for thermal baffling (Gillingham \cite{Gillingham}). However, they interfere with other instrumentation, reducing flexibility. Also the vertical temperature gradient at Dome-C means that a telescope at 30m elevation is much warmer ($\sim-45\dg$ C) than one at ground level, and the relative emission from sky and telescope is very similar to that at a temperate site. Therefore, we have moved to a cold-stop design for the infrared camera, improving the performance and allowing a much more general-purpose telescope design.

\section{Instrumentation}

\begin{enumerate}
\item{There will be a fast readout, zero-noise camera based on an L3Vision 1K x 1K detector, with $\sim 0.02''$ pixels. This will allow optical ``Lucky Imaging'', with $\sim50\%$ of short-exposure frames being diffraction-limited at 800nm in normal conditions. In future, this camera could be given adaptive optics correction to allow diffraction-limited imaging down to 500nm.}

\item{There will be a wide field visible camera sampling
the tip-tilt-corrected natural seeing, over $40' \times 40'$. This might use STA 10K x 10K detectors with $9\mum$ ($0.077''$) pixels , or, if available, we would use Orthogonal Transfer CCDs. The image quality and science potential for this is discussed in Saunders 2008.}

\item{The infrared camera is designed for $1-5\mum$ use. The cold stop is an Offner relay. The detector is a focal plane array of
4k $\times$ 4k HawaiiII-RG with $18\mum$ $0.154''$ pixels, giving diffraction-limited imaging at KLM over a $10.5' \times 10.5'$ field of view. There will be an optional Barlow lens beam expander, giving f/25 imaging, and allowing properly sampled diffraction-limited images at zyJH bands whenever atmospheric conditions allow.
The telescope optics allow an upgrade to much larger detector area, with up to 4 such arrays.}

\item{There will be a mid-infrared camera, using a single Aquarius 1K $\times$ 1K detector, allowing narrow-band imaging at 17 and $21 \mum$, across $\sim0.5 \dg$.}
\end{enumerate}

\section{Image quality}
We have modelled the average $C_N^2$ profile determined by Agabi {\em et al.\/} \cite{Agabi}. There is some residual turbulence in the bounday layer above the telescope, but we can correct for most of this with a fast tip tilt secondary, over the entire field of view. We find that the potential median image quality (diffraction + tip-tilt-corrected seeing) is $<0.3''$ for $0.5-2\mum$ (Figure 4).

The telescope is designed to meet this potential. The specification for image degradation caused by the telescope, including dome and mirror seeing, windshake, guiding errors and optics, is $0.20'' 80\%$ encircled energy diameter, or $0.13''$ FWHM. The expected overall image quality is remarkably uniform over $0.8-2.4\mum$, in the range $0.28-0.3''$.

\begin{figure}
\includegraphics[width=7cm,angle=-90]{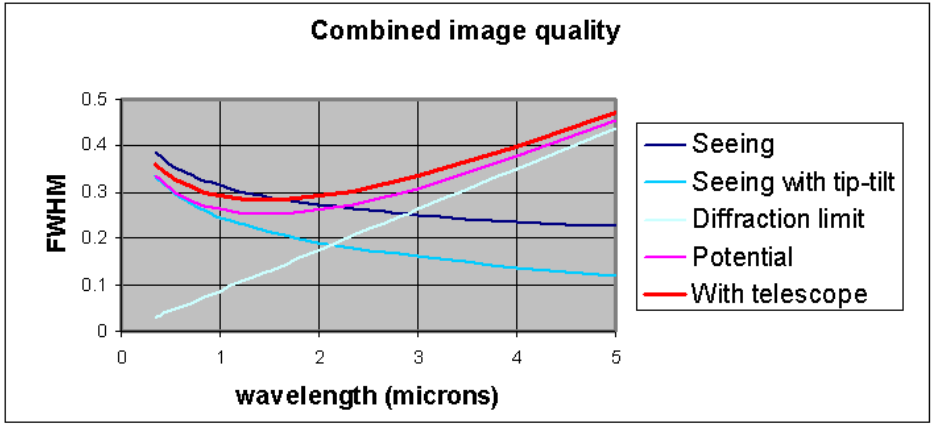}

\caption{Delivered image quality as a function of wavelength, showing the contributions from median seeing, tip-tilt correction, diffraction and telescope. }
\end{figure}

\section{Sensitivity}

The estimated point source sensitivities, based on the image sizes and backgrounds above, are given below. The magnitude limit at $K_{dark}$ is 2-3 magnitudes deeper than VISTA, and deep enough to match the big optical sky surveys such as VST. Other scientific possibilities are outlined in Burton {\em et al.\/} 2005, Burton 2008, and Saunders 2008.

%

\begin{tabular}{lccccccccccc}

Band & $V$ & $I$ & $J$ & $H$ & $K_{dark}$ & $L_{dark}$ & $L$ & $M$ \\
$\lambda (\mum)$ & 0.55& 0.8 & 1.2 & 1.65 & 2.35 & 2.9 & 3.8 & 4.7 \\
$\Delta\lambda (\mum)$  & 0.09 & 0.15 & 0.26  & 0.29 & 0.23 & 0.20& 0.65 & 0.24\\
Sky ($Jy/''^2$)	&6E-6	&2E-5	&5E-4	&1E-3	&1E-4	&2E-3	&2E-1	&5E-1 \\
Sky ($AB/''^2$)	&	22.0 &	20.7		&	17.2 &	16.4	&	18.9	&	15.7	& 10.7 &	 9.7 \\
$AB$, 1hr $5\sigma$ 	&	27.4 	&	26.8	&	25.3 	&	24.8	&	25.6	&	23.5 	&	21.1	&	19.6

\end{tabular}


\end{document}